\documentclass[twocolumn,aps,superscriptaddress,showpacs,nofootinbib,floatfix]{revtex4}

\usepackage{epsfig,bm,feynmf}

\usepackage{graphics}

\usepackage[normalem]{ulem}  % \sout{old text} for strikeout
\usepackage[dvips]{color} % For blue in-text comments and additions

\renewcommand\sout{\bgroup \color{red} \ULdepth=-.5ex \ULset}
\begin{document}

%\preprint{hep-ph/0402135}

%%%%%%%%%%%%%%%%%%%%% Title %%%%%%%%%%%%%%%%%%%%%%

\title{Charmonium production in heavy-ion collisions from the SPS to LHC}

%%%%%%%%%%%%%%%%%%%% Authors %%%%%%%%%%%%%%%%%%%%%

\author{Taesoo Song}\email{songtsoo@yonsei.ac.kr}
\affiliation{Cyclotron Institute, Texas A$\&$M University, College Station, TX 77843-3366, USA}
\author{Kyong Chol Han}\email{khan@comp.tamu.edu}
\affiliation{Cyclotron Institute and Department of Physics and Astronomy, Texas A$\&$M University, College Station, TX 77843-3366, USA}
\author{Che Ming Ko}\email{ko@comp.tamu.edu}
\affiliation{Cyclotron Institute and Department of Physics and Astronomy, Texas A$\&$M University, College Station, TX 77843-3366, USA}

%%%%%%%%%%%%%%%%%%%% Abstract %%%%%%%%%%%%%%%%%%%%%

\begin{abstract}
Using the two-component model that includes charmonium production from both initial nucleon-nucleon hard scattering and regeneration in the produced quark-gluon plasma, we study $J/\psi$ production in heavy-ion collisions at the SPS, RHIC and LHC. For the expansion dynamics of produced hot dense matter, we use a schematic viscous hydrodynamic model with the specific shear viscosity in the quark-gluon plasma and the hadronic matter taken, respectively, to be twice and ten times the lower bound of $1/4\pi$ suggested by the AdS/CFT correspondence. For the initial dissociation and the subsequent thermal decay of charmonia in the hot dense matter, we use the screened Cornell potential to describe the properties of charmonia and perturbative QCD to calculate their dissociation cross sections. Including regeneration of charmonia in the quark-gluon plasma via a kinetic equation with in-medium chamonium decay widths, we obtain a good description of measured $J/\psi$ nuclear modification factors in Pb+Pb collisions at $\sqrt{s_{NN}}=1.73$ GeV at SPS and in Au+Au collisions at $\sqrt{s_{NN}}=200$ GeV at RHIC. A reasonable description of the measured nuclear modification factor of high transverse momenta $J/\psi$ in Pb+Pb collisions at $\sqrt{s_{NN}}=2.76$ TeV at LHC is also obtained.
\end{abstract}

 \keywords{}

\maketitle

%%%%%%%%%%%%%%%%%%%% Text %%%%%%%%%%%%%%%%%%%%%

\section{introduction}

Since $J/\psi$ suppression was first suggested by Matsui and Satz as a signature of quark-gluon plasma (QGP) formation in relativistic heavy-ion collisions~\cite{Matsui:1986dk}, there have been many experimental~\cite{Alessandro:2004ap,Adare:2006ns} and theoretical studies~\cite{Vogt:1999cu,Zhang:2000nc,Zhang:2002ug,Zhao:2007hh,Yan:2006ve} on this very interesting phenomenon; see, e.g., Refs. \cite{Rapp:2008tf,Andronic:2006ky} for a recent review. The original idea of Matsui and Satz was that the color screening in the produced QGP would prohibit the binding of charm and anticharm quarks into the $J/\psi$ and thus suppress its production. However, lattice QCD calculations of the $J/\psi$ spectral function have since shown that the $J/\psi$ can survive above the critical temperature for the QGP phase transition~\cite{Hatsuda04,Datta04}. As a result, the study of $J/\psi$ suppression in relativistic heavy-ion collisions has been changed from being a signature of the QGP to a probe of its properties. Indeed, we have recently shown in a two-component model, which includes $J/\psi$ production from both initial hard nucleon-nucleon scattering and regeneration from charm and anticharm quarks in the produced QGP, that the in-medium effect on $J/\psi$ interactions in the QGP can affect the $J/\psi$ nuclear modification factor and elliptic flow in Au+Au collisions at $\sqrt{s_{NN}}=200$ GeV at the Relativistic Heavy Ion Collider (RHIC) \cite{Song:2010er}. In the present study, we extend this study to $J/\psi$ production in Pb+Pb collisions at the higher energy of $\sqrt{s_{NN}}=2.76$ TeV at the Large Hadron Collider (LHC)~\cite{:2010px,cms} and also at the lower energy of $\sqrt{s_{NN}}=17.3$ GeV at the Super Proton Synchrotron (SPS) \cite{Alessandro:2004ap}. Furthermore, a schematic viscous hydrodynamic model is used to include the effect of viscosity on the expansion dynamics of the produced hot dense matter that was neglected in our previous studies. We find that the two-component model can give a good description of the experimental data from heavy-ion collisions at these different energies.

To make the present paper self contained, we briefly review in Sec.~\ref{two} the two-component model for $J/\psi$ production, in Sec.~\ref{hydrodynamics} the schematic causal viscous hydrodynamical model used in modeling the expansion dynamics of produced hot dense matter, and in Sec.~\ref{properties} the in-medium dissociation temperatures and thermal decay widths of charmonia. Results obtained from our study for the $J/\psi$ nuclear modification factors in heavy-ion collisions at SPS, RHIC and LHC are then presented in Sec. \ref{suppression}. Finally, a summary is given in Sec. \ref{summary}.

\section{The two-component model}\label{two}

The two-component model for $J/\psi$ production in heavy-ion collisions~\cite{Grandchamp:2002wp,Grandchamp:2003uw} includes contributions from both initial hard nucleon-nucleon scattering and regeneration from charm and anticharm quarks in the produced QGP. For initially produced $J/\psi$'s, their number is proportional to the number of binary collisions between nucleons in the two colliding nuclei. Whether these $J/\psi$'s can survive after the collisions depends on many effects from both the initial cold nuclear matter and the final hot partonic and hadronic matters. The cold nuclear matter effects include the Cronin effect of gluon-nucleon scattering before the production of the primordial $J/\psi$ from the gluon-gluon fusion~\cite{Cronin:1974zm}; the shadowing effect due to the modification of the gluon distribution in a heavy nucleus~\cite{Eskola:2009uj}; and the nuclear absorption by the passing nucleons~\cite{Alessandro:2003pi,Lourenco:2008sk,Vogt:2010aa}. In our previous work~\cite{Song:2010fk} on $J/\psi$ production in heavy-ion collisions at RHIC, we have considered only the most important nuclear absorption effect.  In this case, the survival probability of a primordial $J/\psi$ after the nuclear absorption is given by \cite{Kharzeev:1996yx,Ferreiro:2008wc}
\begin{eqnarray}
S_{\rm cnm}({\bf b},{\bf s})=
\frac{1}{T_{AB}({\bf b},{\bf s})}\int dz dz' \rho_A({\bf s},z)\rho_B({\bf b}-{\bf s},z')\nonumber\\
\times {\rm exp}\bigg\{ -(A-1)\int_z^\infty dz_A \rho_A ({\bf s},z_A)\sigma_{\rm abs}\bigg\}~~~~~~~\nonumber\\
\times {\rm exp}\bigg\{ -(B-1)\int_{z'}^\infty dz_B \rho_B
({\bf b}-{\bf s},z_B)\sigma_{\rm abs}\bigg\},
\label{absorption}
\end{eqnarray}
where ${\bf b}$ is the impact parameter and ${\bf s}$ is the transverse vector from the center of nucleus A; $T_{AB}({\bf b},{\bf s})$ is the nuclear overlap function; $\rho_A ({\bf s},z)$ is the density distribution in the nucleus; $\sigma_{\rm abs}$ is the $J/\psi$ absorption cross section by a nucleon. For the latter, it is obtained from p+A collisions and has values of 4.18 and 2.8 mb for the SPS and RHIC, respectively \cite{Alessandro:2004ap,Adare:2007gn}. Presently, there are no p+A data available from the LHC. Since the cross section for $J/\psi$ absorption is expected to decrease with increasing energy~\cite{Lourenco:2008sk}, we consider in the present study the two extreme values of 0 and 2.8 mb to study its effect on the $J/\psi$ yield in heavy-ion collisions at LHC.

Although the shadowing effect has usually been neglected in heavy-ion collisions at SPS and RHIC, this may not be justified at LHC. In the present study, we thus include also the shadowing effect for heavy-ion collisions at LHC using the EPS09 package~\cite{Eskola:2009uj}. The shadowing effect is expressed by the ratio $R_i^A$ of the parton distribution $f_i^A(x,Q)$ in a nucleus to that in a nucleon $f_i^{\rm nucleon}(x,Q)$ multiplied by the mass number $A$ of the nucleus, i.e.,
\begin{eqnarray}
R_i^A(x,Q)=\frac{f_i^A(x,Q)}{A f_i^{\rm nucleon}(x,Q)}, \quad i=q, \bar{q}, g.
\end{eqnarray}
In the above, $x=m_T/\sqrt{s_{NN}}$, with $m_T$ being
the transverse energy of the produced charmonium and $\sqrt{s_{NN}}$ being the center-of-mass energy of colliding nucleons, is the momentum fraction and $Q=m_T$ is the momentum scale. Assuming the shadowing effect is proportional to the path length, we can then express the spatial dependence of $R_i^A$ as \cite{Vogt:2004dh,Lansberg:2005pc,Ferreiro:2008wc}:
\begin{eqnarray}
\frac{R_i^A({\bf s},x,Q)-1}{R_i^A(x,Q)-1}=N\frac{\int dz \rho_A({\bf s},z)}{\int dz \rho_A({\bf 0},z)},
\end{eqnarray}
where $N$ is a normalization factor determined from the condition
\begin{eqnarray}
\frac{1}{A}\int d^2{\bf s}\int dz \rho_A({\bf s},z) R_i^A({\bf s},x,Q)=R_i^A(x,Q).
\end{eqnarray}

\begin{table}[h]
\centering
\begin{tabular}{c| c c c c}
\hline ~~~ & ~SPS~ & ~RHIC~  & ~LHC~  & ~LHC~\\[2pt]
~~~ & ~~~ & ~~~  & ~~~  &$p_T>$6.5 GeV\\[2pt]
\hline production ($\mu$b)~& & & &\\[2pt]
~$d\sigma_{J/\psi}^{pp}/dy$~ & ~0.05~\cite{Andronic:2006ky} & ~0.774~\cite{Adare:2006kf} & ~4.0~ \\[2pt]
~$d\sigma_{c\bar{c}}^{pp}/dy$~ & ~5.7~\cite{Andronic:2006ky} & ~119~\cite{Adare:2010de} & ~615~ &\\[2pt]
\hline feed-down (\%)~&  & & &\\[2pt]
~$f_{\chi_c}$~ & ~25~\cite{Faccioli:2008ir} & ~32~\cite{Adare:2011vq} & ~26.4~\cite{Abe:1997yz} & ~23.5~\cite{Abe:1997yz}\\[2pt]
~$f_{\psi^\prime(2S)}$~ & ~8~\cite{Faccioli:2008ir} & ~9.6~\cite{Adare:2011vq} & ~5.6~\cite{Abe:1997yz} & ~5~\cite{Abe:1997yz}~ \\[2pt]
~$f_b$~& ~~~ & ~~~ & ~11~\cite{Acosta:2004yw} &~21~\cite{Acosta:2004yw}\\[2pt]
\hline
~$R_g^A$ for charm~ & ~~~ & ~~~ & ~0.813~ & ~0.897~\\[2pt]
\hline ~$\tau_0$ (fm/c)~ & ~1.0~ & ~0.9~\cite{Song:2011qa} & ~1.05~\cite{Song:2011qa} \\[2pt]
~$\eta/s$~& ~0.16~ & ~0.16~\cite{Song:2011qa} & ~0.2~\cite{Song:2011qa} &\\[2pt]
\hline
\end{tabular}
\caption{Parameters for $J/\psi$ production and the firecylinder expansion in Pb+Pb collisions at
$\sqrt{s_{NN}}=17.3$ GeV at SPS and at $\sqrt{s_{NN}}=2.76$ TeV at LHC and in Au+Au collisions
at $\sqrt{s_{NN}}=17.3$ GeV at RHIC. $d\sigma_{J/\psi}^{pp}/dy$ and $d\sigma_{c\bar{c}}^{pp}/dy$ are, respectively, the differential $J/\psi$ and $c{\bar c}$ production cross sections in rapidity in p+p collisions; $f_{\chi_c}$, $f_{\psi^\prime(2S)}$, and $f_b$ are, respectively, the fraction of $J/\psi$ production from the decay of $\chi_c$, $\psi^\prime$, and bottom hadrons in p+p collisions; $R_g^A$ is the gluon shadowing effect on charm production; and $\tau_0$ and $\eta/s$ are the thermalization time and the specific viscosity of the produced $QGP$. Also shown in the last column are the parameters for the feed-down contribution to the production of $J/\psi$'s of transverse momentum $p_T > 6.5$ GeV at LHC.} \label{parameters}
\end{table}

The shadowing effect reduces the survival probability of a primordial $J/\psi$ after the nuclear absorption (Eq.(\ref{absorption})) by the factor $R_g^A({\bf s},x,Q)R_g^B({\bf b}-{\bf s},x,Q)$. Taking the momentum scale $Q=4.2~{\rm MeV}$ to be the average $J/\psi$ transverse energy at $\sqrt{s_{NN}}=1.96$ TeV \cite{Acosta:2004yw}, we obtain the value of the ratio $R_g^{\rm pb}(x,Q)$ given in Table \ref{parameters} for charm production in heavy-ion collisions at the LHC.

For the hot partonic and hadronic matter effect, the model includes the dissociation of charmonia in the QGP
of temperatures higher than the dissociation temperature and the thermal decay of survived charmonia through interactions with thermal partons in the expanding hot dense mater. Since the number of produced charm quarks in relativistic heavy-ion collisions is not small, charmonia can also be regenerated from charm and anticharm quarks in the QGP. The effect of thermal dissociation and regeneration of charmonia on the number $N_i$ of charmonium of type $i$ is taken into account via the rate equation \cite{Grandchamp:2003uw}
\begin{eqnarray}
\frac{dN_i}{d\tau}=-\Gamma_i(N_i-N_i^{\rm eq}),
\label{rate}
\end{eqnarray}
where $\tau$ is the longitudinal proper time, while $N_i^{\rm eq}$ and $\Gamma_i$ are, respectively, the equilibrium number and thermal decay width of charmonia and will be discussed in Sec.~\ref{properties}.

Since charm quarks are not expected to be completely thermalized either chemically or kinetically during the expansion of the hot dense matter, the fugacity parameter $\gamma$ and the relaxation factor $R$ are introduced to describe their distributions.  Assuming that the number of charm and anticharm quark pairs does not change during the fireball expansion, the fugacity is obtained from \cite{BraunMunzinger:2000px,Gorenstein:2000ck}
\begin{eqnarray}
N_{c\bar{c}}^{AB}=\bigg\{\frac{1}{2}\gamma n_o\frac{I_1(\gamma n_o V)}{I_0(\gamma n_o V)}+\gamma^2 n_h \bigg\}V,
\label{fugacity}
\end{eqnarray}
where $N_{c\bar{c}}^{AB}$ is the number of $c\bar{c}$ pairs produced in an A+B collision; $n_o$ and $n_h$ are, respectively, the number densities of open- and hidden-charm hadrons in grand canonical ensemble; $V$ is the volume of the hot dense matter; and $I_0$ and $I_1$ are modified Bessel functions resulting from the canonical suppression of charm quarks in heavy-ion collisions~\cite{Gorenstein:2000ck,Ko:2000vp}. For the relaxation factor, it is defined as $R(\tau)=1-\exp[-(\tau-\tau_0)/\tau_{\rm eq}]$ with the relaxation time $\tau_{\rm eq}=3~{\rm fm/c}$ of charm quarks in the QGP taken from Ref. \cite{Zhao:2007hh} and $\tau_0$ being the initial thermalization time.

Since charmonia can only be regenerated in the QGP of temperature below the dissociation temperature $T_i$, the number of equilibrated charmonium of type $i$ in the QGP is
\begin{eqnarray}
N_i^{\rm eq}=\gamma^2 R~ n_i ~f V\theta(T_i-T),
\end{eqnarray}
where $n_i$ is its number density in grandcanonical ensemble;
$f$ is the fraction of QGP in the mixed phase and is 1 in the QGP; and $\theta(T_i-T)$ is the step function.

For the initial charmonium number $N_i$ and the charm quark pair number $N_{c\bar{c}}$, they are obtained from multiplying their respective differential cross sections in rapidity $d\sigma_i^{pp}/dy$ and $d\sigma_{c\bar{c}}^{pp}/dy$ in p+p collisions \cite{Andronic:2006ky,Adare:2006kf,Adare:2010de} by the number of binary collisions $N_{\rm coll}$ in heavy-ion collisions. Since only the $J/\psi$ production cross section at $\sqrt{s}=7$ TeV \cite{Khachatryan:2010yr} has been measured in p+p collisions at LHC, its value at $\sqrt{s_{NN}}=2.76$ TeV is obtained by using a linear function in $\sqrt{s}$ to interpolate from the measured values at $\sqrt{s}=1.96$ TeV by the CDF Collaboration at the Fermi Lab \cite{Acosta:2004yw} to the one at $\sqrt{s}=7$ TeV at LHC. The cross section for $c\bar{c}$ pair production at $\sqrt{s_{NN}}=2.76$ TeV is then determined by assuming that the ratio between the $J/\psi$ and $c\bar{c}$ pair production cross sections is the same as that at RHIC. In Table \ref{parameters}, we list the differential cross sections for $J/\psi$ and $c\bar{c}$ pair production in p+p collision at SPS, RHIC and LHC that are used in the present study.

Since $J/\psi$ production in p+p collisions includes the contribution from the decay of excited charmonium states, the cross section $d\sigma_{J/\psi}^{pp}/dy$ shown in Table \ref{parameters} is the sum of the production cross sections for the $J/\psi$ and its excited states. For p+p collisions at SPS, we use the global average values of the fractions $f_{\chi_c}=$ 25\% from the $\chi_c$ decay and $f_{\psi^\prime(2S)}=$ 8\% from the $\psi'$ decay \cite{Faccioli:2008ir}. The cross sections for $J/\psi$, $\chi_c$ and $\psi^\prime$ production in a p+p collision at the SPS are then given, respectively, by
\begin{eqnarray}
\sigma_{J/\psi}^*&=&0.67 ~\sigma_{J/\psi}\nonumber\\
\sigma_{\chi_c}&=&\frac{0.25 ~\sigma_{J/\psi}}{{\rm Br}(\chi_c \rightarrow J/\psi+X)},\nonumber\\
\sigma_{\psi'}&=&\frac{0.08 ~\sigma_{J/\psi}}{{\rm Br}(\psi' \rightarrow J/\psi+X)},
\label{excited}
\end{eqnarray}
where $\sigma_{J/\psi}^*$ is the cross section for $J/\psi$ production without the feed-down contribution, and `${\rm Br}$' denotes the branching ratio.

For p+p colisions at RHIC, the fractions of $J/\psi$'s from $\chi_c$ and $\psi^\prime (2S)$ decays are taken to be $f_{\chi_c}=$ 32\% and $f_{\psi^\prime}=$ 9.6 \%, respectively, based on recent experimental results by the PHENIX Collaboration \cite{Adare:2011vq}. Since the fractions of $J/\psi$'s from $\chi_c$ and $\psi^\prime(2S)$ decays are not known at LHC, we use the values inferred from $p+{\bar p}$ annihilation at $\sqrt{s_{NN}}=1.96$ TeV by the CDF Collaboration at the Fermi Lab. It was found in these reactions that among promptly produced $J/\psi$'s, about 64\% are directly produced and about 29.7\% from the $\chi_c$ decay, and both are approximately independent of the $J/\psi$ transverse momentum~\cite{Abe:1997yz}. This leads to the fraction of promptly produced $J/\psi$'s from the $\psi^\prime(2S)$ decay to be 6.3\%. Using the experimental result that promptly produced $J/\psi$'s constitute about 89\% of measured $J/\psi$'s~\cite{Acosta:2004yw}, we obtain the fractions  $f_{\chi_c}=26.4\%$ and $f_{\psi^\prime(2S)}=5.6\%$ of measured $J/\psi$'s that are from $\chi_c$ and $\psi^\prime(2S)$ decays, respectively.   Since the fraction of prompt $J/\psi$'s is reduced to 79\% for $J/\psi$ of transverse momentum $p_T>6.5$ GeV~\cite{Acosta:2004yw}, the fractions of measured $J/\psi$'s of $P_T>6.5$ GeV that are from $\chi_c$ and $\psi^\prime(2S)$ decays are reduced to $f_{\chi_c}=23.5\%$ and $f_{\psi^\prime(2S)}=5.0\%$, respectively. Besides the contribution from excited charmonia, the decay of bottom hadrons can also contribute to $J/\psi$ production in high-energy collisions.
This contribution increases significantly with $p_T$ as shown in the experiments by the CDF~\cite{Acosta:2004yw}, CMS~\cite{Khachatryan:2010yr}, LHCb~\cite{Collaboration:2011sp} and ATLAS~\cite{Aaij:2011jh} Collaborations. The fraction is between 5 \% and 10 \% for $p_T < 3$ GeV, depending on the rapidity of the $J/\psi$, then increases to more than 40 \% at $p_T\sim 15$ GeV, and reaches 60-70 \% for $p_T$ above 25 GeV. On the average, about 11 \% of produced $J/\psi$'s are from the decay of bottom hadrons in $p+{\bar p}$ annihilation at $\sqrt{s}=1.96~{\rm TeV}$ at the Fermi Lab \cite{Acosta:2004yw}, and the fraction increases to about 21 \% for $J/\psi$'s of transverse momentum $p_T > 6.5$ GeV \cite{Acosta:2004yw}. These values and other input parameters used in the present study are shown in Table~\ref{parameters}.

\section{A schematic viscous hydrodynamic model}\label{hydrodynamics}

For the expansion dynamics of the hot dense matter formed in relativistic heavy-ion collisions, we describe it by a schematic causal viscous hydrodynamic model recently developed in Ref. \cite{Song:2010fk}. It is based on the assumption that all thermal quantities such as the energy density, temperature, entropy density, and pressure as well as the azimuthal and space-time rapidity components of the shear tensor are uniform along the transverse direction in the hot dense matter.  Assuming the boost-invariance and using the $(\tau, r, \phi, \eta)$ coordinate system
\begin{eqnarray}
\tau&=&\sqrt{t^2-z^2}, ~~\eta=\frac{1}{2}\ln \frac{t+z}{t-z},\nonumber\\
r&=&\sqrt{x^2+y^2}, ~~\phi=\tan^{-1}(y/x),
\end{eqnarray}
then the following equations are obtained from the usual Israel-Stewart viscous hydrodynamic equations:
\begin{eqnarray}
&&\partial_\tau (A\tau \langle T^{\tau \tau}\rangle)=-(p+\pi^\eta_\eta)A,\label{energy7}\\
\nonumber\\
&&\frac{T}{\tau}\partial_\tau (A\tau s \langle \gamma_r\rangle)=-A\bigg\langle\frac{\gamma_r v_r}{r}\bigg\rangle \pi^\phi_\phi-\frac{A\langle \gamma_r\rangle}{\tau}\pi^\eta_\eta\nonumber\\
&&~~~~~~~+\bigg\{\partial_\tau(A\langle \gamma_r\rangle)-\frac{\gamma_R \dot{R}}{R}A\bigg\}(\pi^\phi_\phi+\pi^\eta_\eta),\label{entropy7}\\
\nonumber\\
&&\partial_\tau (A\langle \gamma_r\rangle \pi^\eta_\eta) -\bigg\{\partial_\tau(A\langle\gamma_r\rangle)+2\frac{A\langle\gamma_r\rangle}{\tau} \bigg\}\pi^\eta_\eta\nonumber\\
&&~~~~=-\frac{A}{\tau_\pi}\bigg[\pi^\eta_\eta-2\eta_s\bigg\{\frac{\langle\theta\rangle}{3}-\frac{\langle\gamma_r\rangle}{\tau}\bigg\}\bigg],\label{entropy7}\\
\nonumber\\
&&\partial_\tau(A\langle\gamma_r\rangle~ \pi^\phi_\phi)-\bigg\{\partial_\tau(A\langle\gamma_r\rangle)+2A\bigg\langle\frac{\gamma_r v_r}{r}\bigg\rangle\bigg\}\pi^\phi_\phi\nonumber\\
&&~~~~=-\frac{A}{\tau_\pi}\bigg[ \pi^\phi_\phi-2\eta_s \bigg\{\frac{\langle\theta\rangle}{3}-\bigg\langle\frac{\gamma_r v_r}{r}\bigg\rangle\bigg\}\bigg]\label{shear7b}.
\end{eqnarray}
In the above, $T^{\tau\tau}=(e+P_r)u_\tau^2 -P_r$ is the time-component of the energy-momentum tensor, $\pi^\phi_\phi=r^2\pi^{\phi\phi}$ and $\pi^\eta_\eta=\tau^2\pi^{\eta\eta}$ are, respectively, the azimuthal and  the space-time rapidity component of the shear tensor; $\eta_s$ and $\tau_\pi$ are the shear viscosity of the hot dense matter and the relaxation time for the particle distributions, respectively; $\theta=\frac{1}{\tau}\partial_\tau (\tau \gamma_r)+\frac{1}{r}\partial_r(rv_r \gamma_r)$ with $\gamma_r=1/\sqrt{1-v_r^2}$ in terms of the radial velocity $v_r$; $A=\pi R^2$ with $R$ being the transverse radius of the uniform matter; and $\langle\cdots\rangle$ denotes average over the transverse area. For the radial flow velocity that is a linear function of the radial distance from the center, i.e., $\gamma_r v_r=\gamma_R \dot{R}(r/R)$, where $\dot{R}=\partial R/\partial \tau$ and $\gamma_R=1/\sqrt{1-\dot{R}^2}$, we have $\langle\gamma_r^2\rangle=1+\gamma_R^2 \dot{R}^2/2$, $\langle\gamma_r^2 v_r^2\rangle=\gamma_R^2 \dot{R}^2/2$, $\langle\gamma_r\rangle=2(\gamma_R^3-1)/(3\gamma_R^2 \dot{R}^2)$, and $\langle\gamma_r v_r/r\rangle=\gamma_R \dot{R}/R$. With the energy density $e$ and pressure $p$ related by the equation of state of the matter through its temperature $T$, Eqs.(\ref{energy7})-(\ref{shear7b}) are four simultaneous equations for $T$, $\dot{R}$, $\pi^\phi_\phi$ and $\pi^\eta_\eta$, and can be solved numerically by rewriting them as difference equations.

For the equation of state of the produced dense matter, we use the quasiparticle model with three flavors for the QGP phase \cite{Levai:1997yx,Song:2010ix} and the resonance gas model  for the HG phase. As to the specific shear viscosity $\eta_s/s$, where $s$ is the entropy density, its value in the QGP is taken to be 0.16 for SPS and RHIC, and 0.2 for LHC \cite{Song:2011qa}, while it has the same value of $5/2\pi$ in the HG \cite{Demir:2008tr}. The specific viscosity in the mixed phase is assumed to be their linear combination, i.e., $(\eta/s)_{QGP}f+(\eta/s)_{HG}(1-f)$, where $f$ is the fraction of QGP in the mixed phase. The initial thermalization time is taken to be 1.0 fm/$c$ for SPS, which has usually been used, and 0.9 fm/c and 1.05 fm/c for RHIC and LHC, respectively \cite{Song:2011qa}. Although the initial thermalization time for RHIC is 0.6 fm/c in ideal hydrodynamics \cite{Hirano:2001eu}, the nonzero viscosity generates additional transverse flow \cite{Song:2010fk} and requires a late thermalization to fit the experimental data on $p_T$ spectra and elliptic flows. This is the same reason for the later thermalization at LHC in viscous hydrodynamics.

The initial local temperature of produced matter can be calculated from the equation of state and the local entropy density, which we parameterize as \cite{Song:2010ix,Bozek:2011wa}
\begin{equation}
\frac{ds}{d\eta}=C\left[(1-\alpha)\frac{n_{\rm part}}{2}+\alpha~n_{\rm coll}\right],
\label{entroden}
\end{equation}
with $\alpha=$ 0, 0.11 and 0.15 for SPS, RHIC and LHC, respectively \cite{Antinori:2000ph,Kharzeev:2000ph,Bozek:2011wa}. The number density $n_{\rm part(coll)}$ in Eq. (\ref{entroden}) is defined as $\Delta N_{\rm part(coll)}/(\tau_0\Delta x \Delta y)$, where $\Delta N_{\rm part(coll)}$ is the number of participants (binary collisions) in the volume $\tau_0\Delta x\Delta y$ of the transverse area $\Delta x\Delta y$ and is obtained from the Glauber model with the inelastic nucleon-nucleon cross sections of 30, 42 and 64 mb for SPS, RHIC and LHC, respectively \cite{Back:2004je,Ferreiro:2011rw}. The factor $C$ is determined by fitting the multiplicity of final charged particles after the hydrodynamical evolution to the measured one.

Assuming the same chemical freeze out temperature $T_f=160$ MeV for all charged particles, their pseudorapidity distribution at midrapidity is then \cite{Song:2010er}
\begin{eqnarray}
\frac{dN_{\rm ch}}{d\eta}\bigg|_{y=0}&=&\sum_i \int dp_T \sqrt{1-\frac{m_i^2}{{m_{Ti}}^2}}D_i\frac{dN_i}{dydp_T}\nonumber\\
&=&\frac{\tau}{\pi}\sum_i D_i\int dp_T~ p_T^2 \int_0^R rdr \nonumber\\
&&\times I_0\bigg[\frac{p_T \sinh
\rho}{T_f}\bigg] K_1\bigg[\frac{m_{Ti} \cosh \rho}{T_f}\bigg],
\label{multiplicity}
\end{eqnarray}
where $\rho=\tanh^{-1}(v_r)$. The summation $i$ includes all mesons lighter than 1.5 GeV and all baryons lighter than 2.0 GeV. In including the contribution from the decays of particles, we simply multiply their pseudorapidity distributions by the product $D_i$ of their decay branching ratio and the number of charged particles resulting from the decay. We have thus neglected the difference between the rapidity of the daughter particles and that of the decay particle.  Also, we have used the thermal momentum distributions at chemical freeze out as well as during the expansion of the hot dense matter, thus ignoring the viscous effect on the particle momentum distributions as it is only important for particles of large momenta \cite{Dusling:2009df}. From the multiplicities of charged particles per half participant, $(dN_{\rm ch}/d\eta)/(N_{\rm part}/2)$, which are roughly 2, 4 and 8.4 in central collisions of Pb+Pb at $\sqrt{s_{NN}}=17.3$ GeV at SPS, of Au+Au collisions at $\sqrt{s_{NN}}=200$ GeV at RHIC, and of Pb+Pb at $\sqrt{s_{NN}}=2.76$ TeV at LHC, respectively \cite{Back:2004je,Aamodt:2010cz}, we obtain the corresponding values of 14.6, 18.7 and 27.0 for the parameter $C$ in Eq. (\ref{entroden}).

\begin{figure}[h]
\centerline{
\includegraphics[width=8.5 cm]{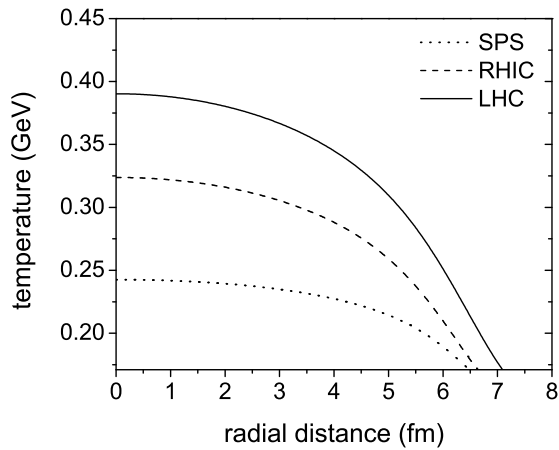}}
\caption{Temperature profiles along the radial direction at initial thermalization time as functions of radial distance in central Pb+Pb collisions at $\sqrt{s_{NN}}=17.3$ GeV at SPS and
$\sqrt{s_{NN}}=2.76$ TeV at LHC, and in central Au+Au collisions at $\sqrt{s_{NN}}=200$ GeV at RHIC from viscous hydrodynamics.}\label{temperaturea}
\end{figure}

\begin{figure}[h]
\centerline{
\includegraphics[width=8.5 cm]{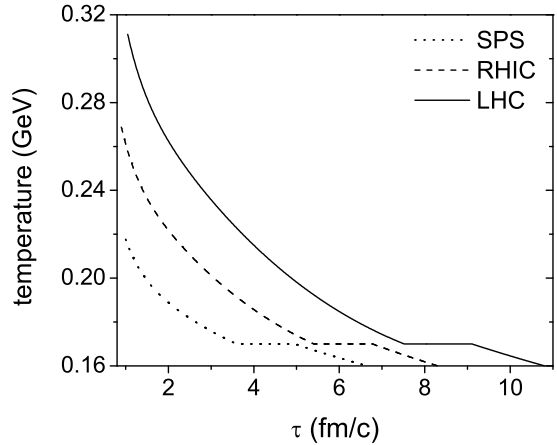}}
\caption{Average temperatures of firecylinder as functions of time in central Pb+Pb collisions at $\sqrt{s_{NN}}=17.3$ GeV at SPS and $\sqrt{s_{NN}}=2.76$ TeV at LHC, and in central Au+Au collisions at $\sqrt{s_{NN}}=200$ GeV at RHIC from the viscous hydrodynamics.}
\label{temperatureb}
\end{figure}

In Fig. \ref{temperaturea}, we show the temperature profile along radial direction at initial thermalization time in heavy-ion collisions at SPS, RHIC, and LHC from the viscous hydrodynamics. Defining the firecylinder as the region where the initial temperature is above $T_c=170~{\rm MeV}$, its transverse radius in the case of viscous hydrodynamics has values of 6.5, 6.6 and 7.1 fm in central collisions at SPS, RHIC and LHC, respectively. The time evolution of the average temperature of the firecylinder determined from the schematic hydrodynamic model is shown in Fig. \ref{temperatureb}. The initial average temperatures at SPS and RHIC are 218 and 269 MeV, respectively, and are consistent with those extracted from the experimental data on dileptons at SPS \cite{Collaboration:2010xu} and on direct photons at RHIC \cite{:2008fqa}. The predicted initial average temperature in heavy-ion collisions at LHC is 311 MeV.

For non-central heavy-ion collisions where the initial geometry of the transverse area is an ellipse, the schematic viscous hydrodynamic model described here needs to be extended. For simplicity, the present model is used by taking the circular transverse area to be the same as that of the ellipse as in Ref.~\cite{Song:2010ix} based on a parameterized firecylinder model.

\section{thermal properties of charmonia}\label{properties}

To describe the properties of charmonia in QGP, we need the potential between heavy quark and its antiquark at finite temperature. Although some information on this can be obtained from the lattice gauge theory \cite{Kaczmarek:2003dp,Wong:2004zr} or from the static QCD \cite{Brambilla:2008cx,Brambilla:2010vq}, we use in the present study the extended Cornell model that includes the Debye screening effect on color charges \cite{Karsch:1987pv}. The Cornell model \cite{Eichten:1979ms} was devised to imitate the asymptotic freedom and confinement of the QCD interaction with a Coulomb-like potential for short distance and a linear potential for long distance. In the QGP, the linear potential becomes weaker due to the Debye screening between color charges, leading to the screened Cornell potential \cite{Karsch:1987pv}
\begin{eqnarray}
V(r,T)=\frac{\sigma}{\mu(T)}\bigg[1-e^{-\mu(T) r}\bigg]-\frac{\alpha}{r}e^{-\mu(T) r}
\label{Cornell}
\end{eqnarray}
with $\sigma=0.192~{\rm GeV^2}$ and $\alpha=0.471$.
The screening mass $\mu(T)$ depends on temperature and is given in thermal pQCD by
\begin{eqnarray}
\mu(T)=\sqrt{\frac{N_c}{3}+\frac{N_f}{6}}~gT,
\label{screening}
\end{eqnarray}
where $N_c$ is the number of colors, $N_f$ is the number of light quark flavors, and $g$ is the QCD coupling constant. In the limit of $\mu \rightarrow 0$, we recover the original Cornell potential.

\begin{figure}[h]
\centerline{
\includegraphics[width=9 cm]{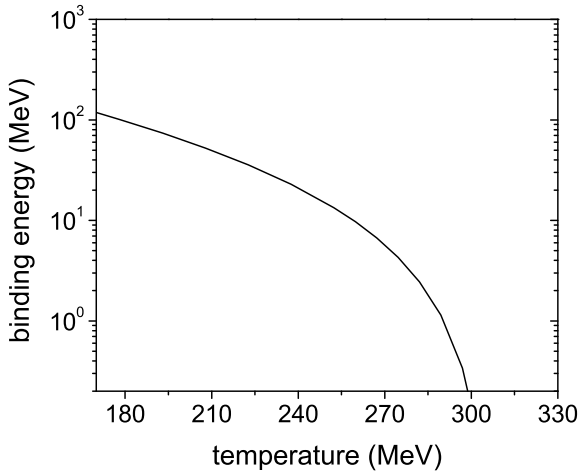}}
\caption{Binding energy of $J/\psi$ in the QGP as a function of temperature for the QCD coupling constant
$g=1.87$.}
\label{bindingE}
\end{figure}

\begin{figure}[h]
\centerline{
\includegraphics[width=9 cm]{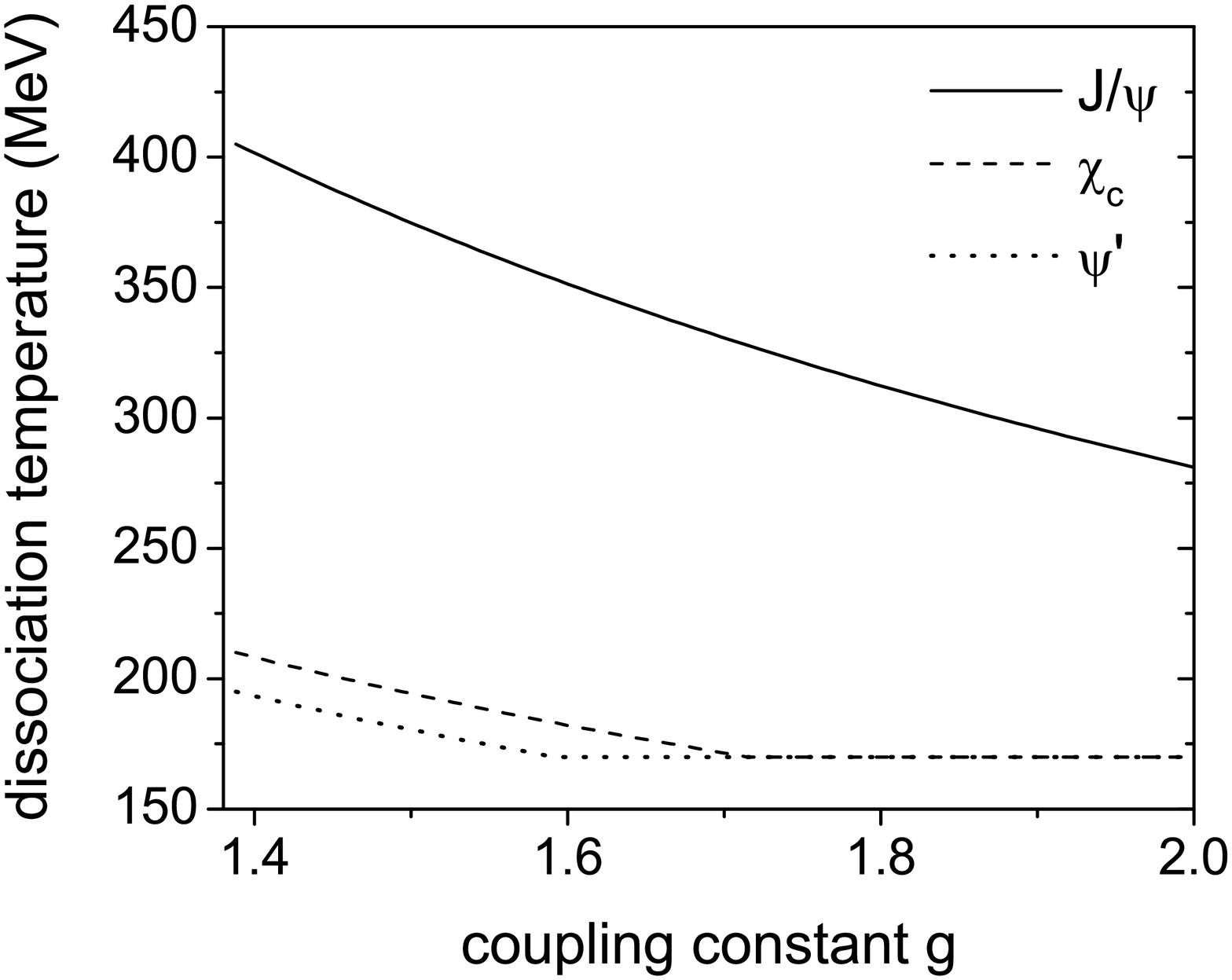}}
\caption{Charmonium dissociation temperatures in the QGP as functions of the QCD coupling constant $g$.}
\label{disso}
\end{figure}

The wavefunctions and binding energies of charmonia in the QGP are obtained by solving the Schr\"odinger equation with the screened Cornell potential. With the binding energy $\varepsilon_0$ defined as \cite{Karsch:1987pv}
\begin{eqnarray}
\varepsilon_0=2m_c+\frac{\sigma}{\mu}-E,
\end{eqnarray}
where the charm quark mass is taken to be $m_c=1.32~{\rm GeV}$ and $E$ is the eigenvalue of the Schr\"odinger equation, we show in Fig.~\ref{bindingE} the binding energy of $J/\psi$ as a function of temperature for the case of $g=1.87$.
It is seen that the $J/\psi$ becomes unbound or dissociated in the QGP for temperatures above $\sim 300$ MeV. As indicated by Eq. (\ref{screening}), the $J/\psi$ dissociation temperature decreases as the QCD coupling constant $g$ increases. This is shown in Fig.~\ref{disso} not only for the $J/\psi$ but also for its excited states $\chi_c$ and $\psi^\prime$. In obtaining the dissociation temperatures for $\chi_c$ and $\psi^\prime$, we have assumed that they are always above the critical temperature $T_c=170~{\rm MeV}$ even for large $g$. We note that the screening mass $\mu$ is nonzero in QCD vacuum but has a value of 180 MeV~\cite{Karsch:1987pv}. In this case, the binding energy of $J/\psi$ in the vacuum is 600$\sim$700 MeV.

Although the charmonium can be formed in the QGP at high temperature, it can still be dissociated by scattering with thermal partons. In the leading order (LO) pQCD, the charmonium breaks up by absorbing a thermal gluon, while in the next-to-leading order (NLO) the dissociation is induced either by a quark or a gluon, and their invariant matrix elements are given, respectively, by \cite{Park:2007zza}
\begin{eqnarray}
\overline{|\mathcal{M}|}_{\rm LO}^2=\frac{2}{3N_c}g^2m_C^2m_\Phi
(2k_{10}^2+m_G^2) {\Big|\frac{\partial \psi({\bf
p})}{\partial {\bf p}}\Big|}^2,
\label{LO}
\end{eqnarray}
\begin{eqnarray}
\overline{|\mathcal{M}|}_{\rm qNLO}^2=\frac{4}{3} g^4 m_C^2 m_\Phi
{\Big|\frac{\partial \psi({\bf p})}{\partial {\bf p}}\Big|}^2
\bigg\{-\frac{1}{2}+\frac{k_{10}^2+k_{20}^2}{2 k_1 \cdot k_2}\bigg\},\nonumber\\
\label{qNLO}
\end{eqnarray}
\begin{eqnarray}
\overline{|\mathcal{M}|}_{\rm gNLO}^2=\frac{4}{3} g^4 m_C^2 m_\Phi
{\Big|\frac{\partial \psi({\bf p})}{\partial {\bf p}}\Big|}^2
\Bigg\{-4+\frac{k_1 \cdot k_2}{k_{10}k_{20}}\nonumber \\
+\frac{2k_{10}}{k_{20}}+\frac{2k_{20}}{k_{10}}
-\frac{k_{20}^2}{k_{10}^2}-\frac{k_{10}^2}{k_{20}^2} +\frac{2}{k_1
\cdot k_2}~~~~~\nonumber\\
\times\bigg[
\frac{(k_{10}^2+k_{20}^2)^2}{k_{10}k_{20}} -2 k_{10}^2-2
k_{20}^2+k_{10}k_{20}\bigg] \Bigg\}.
\label{gNLO}
\end{eqnarray}
In the above, $k_1$ and $k_2$ are, respectively, the momenta of incoming and outgoing thermal partons; $\psi({\bf p})$ is the wavefunction of charmonium with ${\bf p}=({\bf k}_1-{\bf k}_2)/2$; $N_c$ is the number of colors; $m_G$ is the mass of thermal gluon and can be extracted from the lattice QCD \cite{Levai:1997yx}; $m_\Phi$ is the mass of charmonium in the QGP; and $m_C\equiv m_c+\sigma/2\mu$ is the mass of the constituent charm quark. With the screening mass $\mu=0.18~{\rm GeV}$ in the vacuum~\cite{Karsch:1987pv}, the latter has a value of $m_C=1.85~{\rm GeV}$ in the vacuum and is similar to the mass of $D$ meson. The dissociation cross sections of charmonia are then obtained by integrating Eq. (\ref{LO})-(\ref{gNLO}) over the phase space.

The same pQCD formula can be used for charmonium dissociation by partons inside hadrons in the HG. It was found, however, that the charmonium is not heavy enough for pQCD to be applicable \cite{Song:2005yd}. In the present study, we thus take the cross section for charmonium dissociation by a hadron to be proportional to its squared radius as in Ref. \cite{Song:2010ix} or given by that from a phenomenological hadronic Lagrangian \cite{Lin:1999ad,Lin:2000ke}. We note that the effect of charmonium dissociation in the HG is negligible compared to that in the QGP due to the much smaller thermal decay width \cite{Grandchamp:2002wp,Song:2010ix}.

In terms of its dissociation cross section $\sigma_i^{\rm diss}$, the thermal decay width of a charmonium is given by
\begin{eqnarray}
\Gamma(T)&=&\sum_i \int\frac{d^3k}{(2\pi)^3}v_{\rm rel}(k)n_i(k,T) \sigma_i^{\rm diss}(k,T),
\label{width}
\end{eqnarray}
where $i$ denotes the quarks and gluons in the QGP, and the baryons and mesons in the HG; $n_i$ is the number density of particle $i$ in grand canonical ensemble; and $v_{\rm rel}$ is the relative velocity between charmonium and the particle. For the thermal width in the mixed phase, it is taken to be a linear combination of those in the QGP and the HG as following:
\begin{equation}
\Gamma(T_c)=f~\Gamma^{\rm QGP}(T_c)+(1-f)\Gamma^{\rm HG}(T_c),
\label{mixed}
\end{equation}
where $f$ is the fraction of QGP in the mixed phase.

\begin{figure}[h]
\centerline{
\includegraphics[width=8.5 cm]{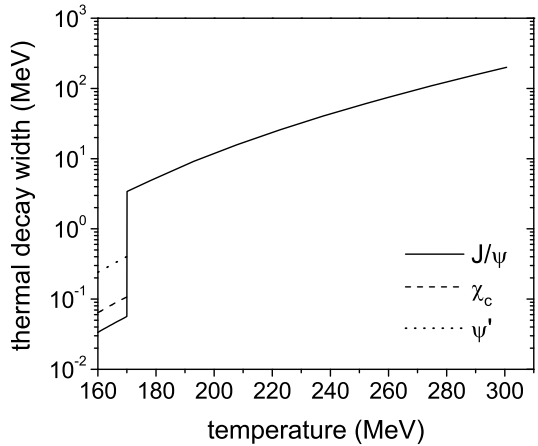}}
\caption{Thermal decay widths of charmonia in the QGP as a function of temperature for the QCD coupling constant $g=1.87$.}
\label{widths}
\end{figure}

The thermal decay widths of charmonia also depend both on the QCD coupling constant and the temperature of QGP. In Fig. \ref{widths}, they are shown as functions of temperature for $g=1.87$. It is seen that the thermal decay width of $J/\psi$ diverges at the dissociation temperature $T=300~{\rm MeV}$, while those of $\chi_c$ and of $\psi^\prime$ become divergent at the critical temperature $T_c=170~{\rm MeV}$. An infinitely large thermal decay width implies that the particles instantly reach their maximally allowed equilibrium value $N_i^{\rm eq}$. Therefore, the $J/\psi$ abundance is not expected to reach this value at $T_c$, in contrast to that of the $\chi_c$ and $\psi^\prime$. We note that the value $g=1.87$ is slightly larger than that used in our previous studies based on a schematic firecylinder model as a result of the viscous effect that is included in the present study.

\section{Results}\label{suppression}

Using the above described two-component model based on the schematic viscous hydrodynamics and taking into account the in-medium effects on charmonia, we can calculate the nuclear modification factor $R_{AA}$ of $J/\psi$ in heavy ion collisions according to
\begin{eqnarray}
R_{AA}&=&(1-f_{\chi_c}-f_{\psi^\prime(2S)}-f_b)R_{\rm pri}+f_b R_b+R_{\rm reg},\nonumber\\
\label{Raa-highpt}
\end{eqnarray}
where $R_{\rm pri}$, $R_b$, and $R_{\rm reg}$ are the nuclear modification factors for $J/\psi$'s that are produced from primordial hard nucleon-nucleon scattering, the decay of bottom hadrons, and the regeneration in the QGP, respectively.
In writing the above expression, we have used the fact that all primordial $\chi_c$ and $\psi^\prime$ are dissociated
above the critical temperature $T_C$. For $R_{\rm pri}$, it is calculated according to \cite{Song:2010ix}:
\begin{eqnarray}
R_{\rm pri}(\vec{b})=\int d^2 {\bf s}~ S_{\rm cnm}({\bf b},{\bf s}){\rm exp}\bigg\{-\int_{\tau_0}^{\tau_f} \Gamma_{J/\psi}d\tau\bigg\},\nonumber\\
\end{eqnarray}
where $\tau_f$ is the freeze-out proper time.  For $R_b$, it is taken to be one as a result of the expected conservation of total bottom and antibottom numbers. As to $R_{\rm reg}$, it is calculated from the ratio of the number of $J/\psi$'s obtained from solving Eq.(\ref{rate}) to the number of $J/\psi$'s from p+p collisions at same energy multiplied by the number of binary collisions in A+A collisions.

\subsection{Nuclear modification factor of $J/\psi$ at SPS and RHIC}

\begin{figure}[h]
\centerline{
\includegraphics[width=7.5 cm]{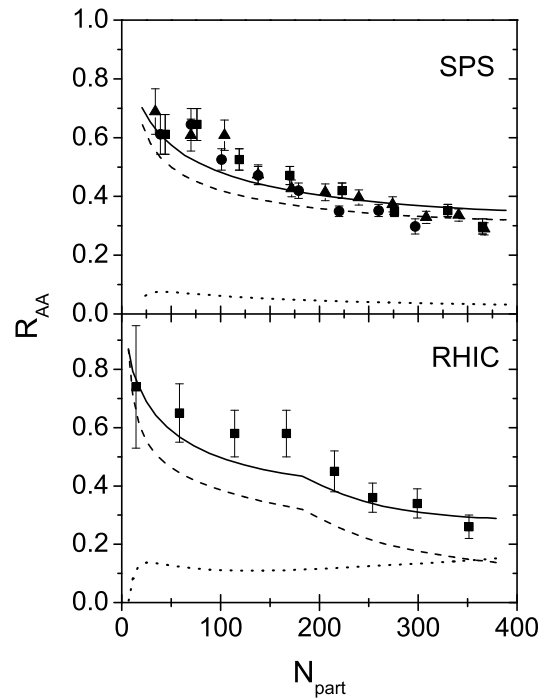}}
\caption{Nuclear modification factor $R_{AA}$ of $J/\psi$ (solid line) as a function of the participant number $N_{\rm part}$ in Pb+Pb collisions at $\sqrt{s_{NN}}=17.3$ GeV at SPS (upper panel) and in Au+Au collisions at $\sqrt{s_{NN}}=200$ GeV at RHIC (lower panel). Dashed and dotted lines represent, respectively, the contributions to $J/\psi$ production from primordial hard nucleon-nucleon scattering and regeneration in the QGP. Experimental data are from Refs. \cite{Alessandro:2004ap,Adare:2006ns}.}
\label{Raa}
\end{figure}

In Fig. \ref{Raa}, we show the nuclear modification factor $R_{AA}$ of $J/\psi$ as a function of the participant number in Pb+Pb collisions at $\sqrt{s_{NN}}=17.3$ GeV at SPS and in Au+Au collisions at $\sqrt{s_{NN}}=200$ GeV at RHIC. These results are obtained with the QCD coupling constant $g=1.87$, which gives a good description of the experimental data as shown by solid lines in the upper and lower panels. It is seen that the $R_{AA}$ of $J/\psi$ becomes smaller as the number of participants in the collision increases. Also shown in Fig. \ref{Raa} are results from the primordial (dashed lines) and the regenerated $J/\psi$ in the QGP (dotted lines), and they clearly indicate that the contribution from the primordial $J/\psi$ decreases and that from the regenerated ones increases as the collision energy increases.

\subsection{Nuclear modification factor of $J/\psi$ at LHC}

\begin{figure}[h]
\centerline{
\includegraphics[width=7.5 cm]{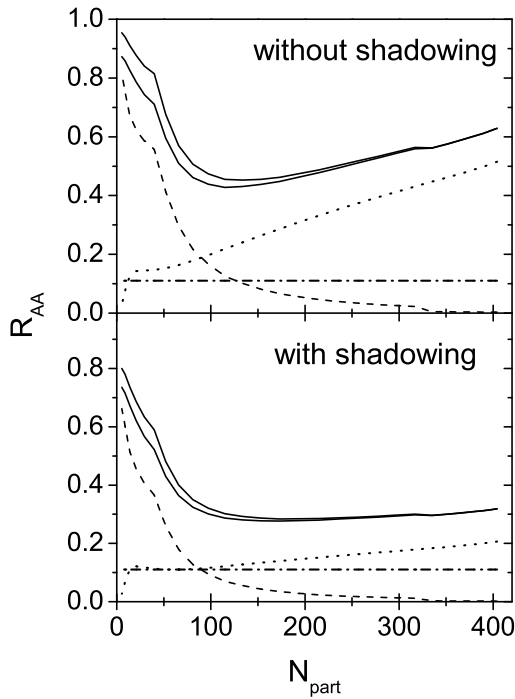}}
\caption{Nuclear modification factor $R_{AA}$ of $J/\psi$ as a function of the participant number $N_{\rm part}$ without (upper panel) and with (lower panel) the shadowing effect in Pb+Pb collisions at $\sqrt{s_{NN}}=2.76$ TeV at LHC. Upper and lower solid lines are the $R_{AA}$ of $J/\psi$ for the nuclear absorption cross sections of $\sigma_{\rm abs}=$0 and 2.8 mb, respectively. Dashed, dotted, dot-dashed lines denote, respectively, the contributions to $J/\psi$ production from primordial hard nucleon-nucleon scattering, regeneration in the QGP, and decay of bottom hadrons.}
\label{Raalhc}
\end{figure}

In Fig. \ref{Raalhc}, we show the $R_{AA}$ of $J/\psi$ in Pb+Pb collisions at $\sqrt{s_{NN}}=2.76$ GeV at LHC with (lower panel) and without the shadowing effect (upper panel). It is seen that the shadowing effect suppresses the production of charm pairs and consequently the regeneration of $J/\psi$. In obtaining these results, we have included the contribution to $J/\psi$ production from the decay of bottom hadrons, which becomes non-negligible at LHC~\cite{Collaboration:2011sp,Aaij:2011jh}, by assuming that the $R_{AA}$ of $J/\psi$ from the decay of bottom hadrons is independent of the centrality as indicated by the measured data from the CMS Collaboration \cite{cms} and shown by the dash-dotted line in Fig. \ref{Raalhc} as a function of the participant number. This contribution is comparable to that from the regenerated $J/\psi$ (dotted lines) in peripheral collisions and more important than the primordial ones (dashed lines) in more central collisions. The upper and lower solid lines are the final $R_{AA}$ of $J/\psi$ obtained with the nuclear absorption cross section of 0 and 2.8 mb, respectively. It is seen that the difference between the results obtained with and without the nuclear absorption is mainly in collisions of small number of participants as the primordial $J/\psi$s are mostly dissolved in central and semi-central collisions.

\begin{figure}[h]
\centerline{
\includegraphics[width=8.5 cm]{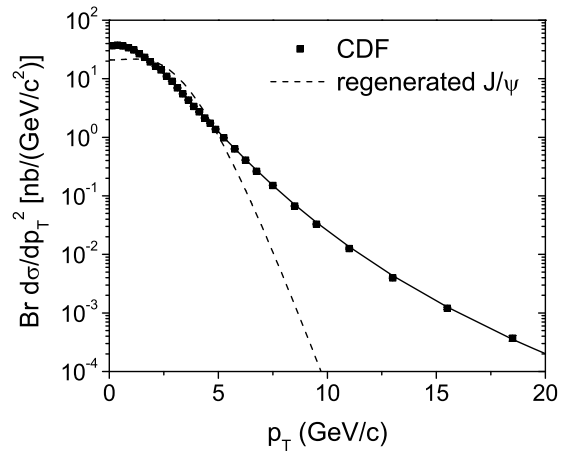}}
\caption{Transverse momentum $p_T$ spectrums of $J/\psi$ from $p+{\bar p}$ annihilation at $\sqrt{s_{NN}}=1.96$ TeV by the CDF Collaboration at Fermi Lab (filled squares and solid line) \cite{Acosta:2004yw} and of regenerated $J/\psi$ (dashed line) in central Pb+Pb collisions at $\sqrt{s_{NN}}=2.76$ TeV at LHC.}
\label{spectrum}
\end{figure}

To compare the results from our model with the experimental data from LHC~\cite{:2010px,cms}, which have a transverse momentum cut $p_T>6.5$ GeV for the measured $J/\psi$, we note that the fraction of produced $J/\psi$'s with transverse momentum larger than 6.5 GeV from the decay of bottom hadrons is about 21\% in $p+{\bar p}$ annihilation at 1.96 TeV from the CDF Collaboration at the Fermi Lab \cite{Acosta:2004yw}. Parameterizing the latter by $~[1+(p_T/4.1{\rm GeV})^2]^{-3.8}$ as shown by the solid line in Fig.~\ref{spectrum}, we obtain that the fraction of $J/\psi$'s with transverse momentum larger than 6.5 GeV is 3\%. This is significantly larger than that from the regeneration contribution in Pb+Pb collisions, which is only 0.17\%, as shown by the dashed line that is obtained from the two-component model but is arbitrarily normalized. It was first pointed out in Ref. \cite{Zhao:2011cv} that limiting the $J/\psi$ transverse momentum to 6.5 GeV suppresses the contribution from the regenerated $J/\psi$. For $J/\psi$'s of high transverse momenta, their nuclear modification factor $R_{AA}$ can thus be calculated by multiplying the last term in Eq.(\ref{Raa-highpt}) by the percentage of regenerated $J/\psi$'s with transverse momentum larger than 6.5 GeV divided by the percentage of primordial $J/\psi$'s with the same range of transverse momenta, which is 0.12 in central Pb+Pb collisions at $\sqrt{s_{NN}}=2.76$ TeV.

\begin{figure}[h]
\centerline{
\includegraphics[width=7.5 cm]{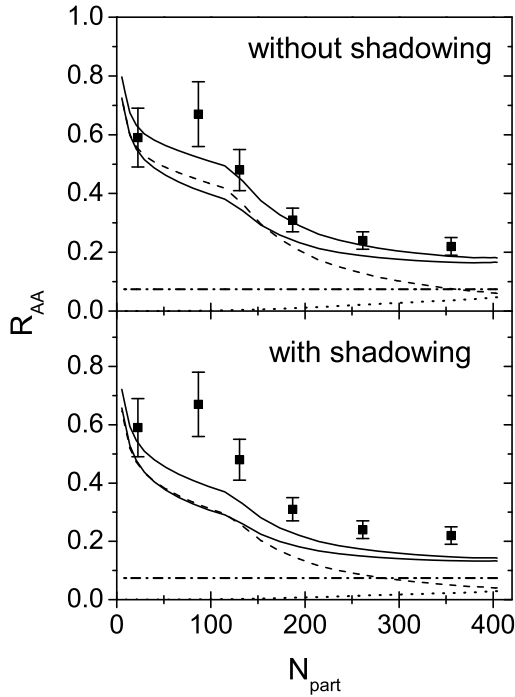}}
\caption{Nuclear modification factor $R_{AA}$ of $J/\psi$ with transverse momentum larger than 6.5 GeV versus the number of participants without (upper panel) and with (lower panel) the shadowing effect in Pb+Pb collisions at $\sqrt{s_{NN}}=2.76$ TeV at LHC. Dashed, dotted and dot-dashed lines represent, respectively, the contributions to $J/\psi$ production from primordial hard nucleon-nucleon scattering, regeneration from QGP, and decay of bottom hadrons, and the solid line is the sum of them, without nuclear absorption. Upper and lower solid lines are the $R_{AA}$ obtained with the nuclear absorption cross section of 0 and 28 mb, respectively.}
\label{LHCa}
\end{figure}

Since it takes time for an initially produced $c\bar{c}$ pair to form a charmonium, which depends on the charmonium radius and the relative velocity between charm and anticharm quark \cite{Blaizot:1988ec,Karsch:1987zw},
thermal dissociation of charmonia is thus delayed until charmonia are formed. Since the $J/\psi$ formation time increases with its transverse momentum as a result of time dilation, this effect becomes more important for $J/\psi$'s of high transverse momenta that are measured in experiments at LHC. In this study, we treat the formation time as a free parameter to fit the experimental data. Using the formation time of 0.5 fm/$c$, which corresponds to 1.4 fm/$c$ in the firecylinder frame based on the average of the $J/\psi$ transverse momenta that are above 6.5 GeV, our results for the the nuclear modification factor $R_{AA}$ of $J/\psi$ with transverse momentum larger than 6.5 GeV as a function of the participant number are shown in Fig. \ref{LHCa} without (upper panel) and with (lower panel) the shadowing effect as well as without (upper solid curve) and with (lower solid curve) the nuclear absorption effect. It is seen that the results obtained without the shadowing and the nuclear absorption effect describe reasonably the recent experimental results from the CMS Collaboration at LHC \cite{cms} shown by solid squares. Also, it is interesting to see that the shoulder structure around $N_{\rm part}=100$ in the measured $R_{AA}$ at LHC is roughly reproduced by our model. As suggested in our previous study on the $J/\psi$ $R_{AA}$ in Au+Au collisions at RHIC, the sudden drop in its value at certain value of $N_{\rm part}$ reflects the maximum temperature of the formed QGP that is above the dissociation temperature of $J/\psi$, because the survival probability of $J/\psi$ is discontinuous at its dissociation temperature. Moreover, the fact that the shoulder seen at LHC occurs at a smaller number of participants than the value $N_{\rm part}=190$ at RHIC is consistent with the expectation that the maximum temperature of the QGP formed at LHC that is above the dissociation temperature of $J/\psi$ happens in more peripheral collisions than at RHIC. We note that the CMS Collaboration has also measured the fraction of $J/\psi$ from the decay of bottom hadrons in Pb+Pb collisions at $\sqrt{s_{NN}}=2.76$ TeV and found that its nuclear modification factor is about 0.37, which corresponds to $R_b$ in Eq. (\ref{Raa-highpt}), and is almost independent of the centrality \cite{cms}.

\begin{figure}[h]
\centerline{
\includegraphics[width=10.0 cm]{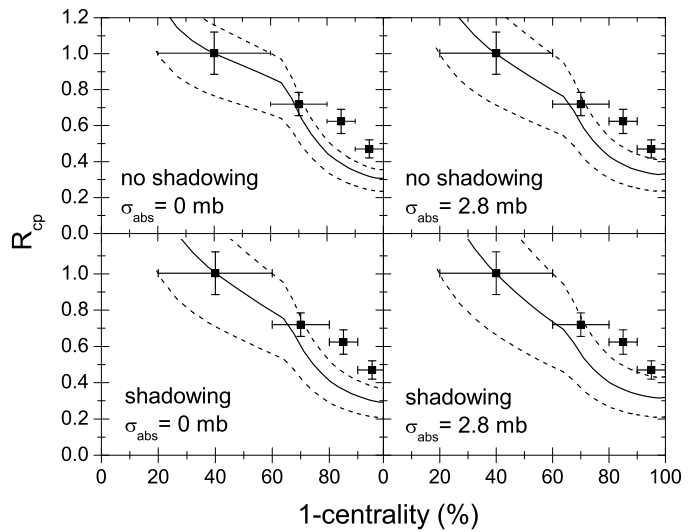}}
\caption{Ratio $R_{\rm cp}$ of the $R_{AA}$ of $J/\psi$
with transverse momentum larger than 6.5 GeV in a given centrality to that
in the peripheral collision versus the centrality of Pb+Pb collisions
at $\sqrt{s_{NN}}=2.76$ TeV at LHC. Experimental data shown by solid squares are from
Ref. \cite{:2010px}.}
\label{LHCb}
\end{figure}

Our results can further be compared with the experimental data from the ATLAS collaboration at LHC \cite{:2010px} on the centrality dependence of the ratio $R_{\rm cp}$ of the $R_{AA}$ of $J/\psi$ in a collision of certain centrality to that in the peripheral collision. For this purpose, we determine the centrality of a collision using the Glauber model as follows \cite{Miller:2007ri}:
\begin{eqnarray}
{\rm Centrality}(b)=\frac{\sigma_{\rm inel}^{AB}(b)}{\sigma_{\rm total~inel}^{AB}}~~~~~~~~~~~~~~~~~~~~~~~~~\nonumber\\
=\frac{\int^b_0 2\pi b^\prime db^\prime \bigg\{1-\bigg[1-T_{AB}(b^\prime)\sigma_{\rm inel}^{NN}\bigg]^{AB}\bigg\}}{\int^\infty_0 2\pi b^\prime db^\prime \bigg\{1-\bigg[1-T_{AB}(b^\prime)\sigma_{\rm inel}^{NN}\bigg]^{AB}\bigg\}},
\end{eqnarray}
where the numerator is the inelastic cross section of nuclei A and B with the impact parameters between 0 and b, and the denominator is the total inelastic cross section of the two nuclei; and $\sigma^{NN}_{\rm inel}$ is the inelastic cross section of a p+p collision at the same collision energy. In Fig. \ref{LHCb}, we show the calculated centrality dependence of $R_{cp}$ in Pb+Pb collisions at $\sqrt{s_{NN}}=2.76$ TeV at LHC with the uncertainty of the reference point, i.e., the $R_{AA}$ of $J/\psi$ in the peripheral collision, shown as dashed lines. It is seen that results from our model calculations can reproduce the measured $R_{\rm cp}$ of $J/\psi$'s with high transverse momenta, and the shadowing and the nuclear absorption effect do not make significant difference in the $R_{\rm cp}$ of $J/\psi$.

\section{summary}\label{summary}

Modeling the evolution of the hot dense matter produced in relativistic heavy-ion collisions by a schematic viscous hydrodynamics, we have extended the two-component model, that was previously used to describe $J/\psi$ production in heavy-ion collisions at RHIC, to those at SPS and LHC. As in our previous studies, we have included the effect due to absorption by the cold nuclear matter on the primordially produced charmonia from initial nucleon-nucleon hard scattering, the dissociation of survived charmonia in the produced hot dense matter, and the regeneration of chamronia from charm and anticharm quarks in the quark-gluon plasma. For heavy-ion collisions at LHC, we have further included the shadowing effect in the initial cold nuclei. We have also taken into account the medium effects on the properties of the charmonia and their dissociation cross sections by using the screened Cornell potential model and the NLO pQCD. With the same quasiparticle model for the equation of state of the QGP and the resonance gas model for that of the HG as used before, we have obtained a lower initial temperature than in our previous study to reach the same final entropy density as a result of the finite viscosity. Consequently, a slightly larger QCD coupling constant was needed to reproduce the measured centrality dependence of the nuclear modification factor at RHIC. The calculated nuclear modification factor for heavy-ion collisions at the SPS was found to agree with the measured value as well. For both the SPS and RHIC, the contribution from the primordial charmonia was found to dominate, although the contribution from the regenerated ones increases from the SPS to RHIC. For heavy-ion collisions at LHC, the regenerated charmonia becomes most important in semi-central to central collisions as a result of the larger number of charm and anticharm quark pairs produced in higher energy collisions. Since the available experimental data from the LHC are for $J/\psi$'s of transverse momentum $p_T > 6.5$ GeV, we have further considered the contribution of $J/\psi$ production from the decay of bottom hadrons as its effect increases with increasing $J/\psi$ transverse momentum and the effect due to the formation time of the $J/\psi$. A reasonable agreement with the preliminary experimental data has been obtained if the shadowing and the nuclear absorption effect is absent. However, a definitive conclusion can only be made after more refined experimental data becomes available. Furthermore, we have found the similar trend in the centrality dependence of the $R_{AA}$ of $J/\psi$ at both RHIC and LHC that it decreases monotonously in peripheral collisions and then drops at a certain centrality as a result of the onset of an initial QGP temperature higher than the $J/\psi$ dissociation temperature. Moreover, this takes place in less central collisions at LHC than at RHIC, indicating that the initial temperature at the same centrality is higher at LHC than at RHIC.

\section*{Acknowledgements}
This work was supported in part by the U.S. National Science
Foundation under Grant Nos. PHY-0758115 and PHY-1068572, the US Department of Energy
under Contract No. DE-FG02-10ER41682, and the Welch Foundation under
Grant No. A-1358.

%%%%%%%%%%%%%%%%%%%%%%%%%%%%%%%%%%%%%%%%%% reference %%%%%%%%%%%%%%%%%%%%%%%%%%%%%%%%%%%%%%%%%%%%%%%%%%%%%%%%%

\end{document}